\documentclass[10pt,aps,prl,twocolumn,superscriptaddress]{revtex4-2}
\usepackage[linkcolor = blue, citecolor = blue, urlcolor = blue, colorlinks = true]{hyperref}
\usepackage[usenames,dvipsnames,x11names]{xcolor}
\usepackage[normalem]{ulem}
\usepackage{graphicx}
\usepackage{listings}
\usepackage{stmaryrd}
\usepackage{amssymb}
\usepackage{amsmath}
\usepackage{gensymb}
\usepackage{float}
\usepackage{bbold}
\usepackage{bm}

\newcommand{\Sp}{\bm{\sigma}^{({\rm p})}}

\newcommand{\Sa}{\bm{\sigma}^{({\rm a})}}

\hypersetup{
    colorlinks=true,
    linkcolor=blue,
    citecolor=red}

\begin{document}

\title{Quasi long-ranged order in two-dimensional active liquid crystals}

\author{Livio Nicola Carenza}
\affiliation{Instituut-Lorentz, Universiteit Leiden, P.O. Box 9506, 2300 RA Leiden, The Netherlands}
\affiliation{Department of Physics, Ko{\c c} University, Rumelifeneri Yolu, 34450 Sarıyer, Istanbul, T{\"u}rkiye}
\author{Josep-Maria Armengol-Collado}
\affiliation{Instituut-Lorentz, Universiteit Leiden, P.O. Box 9506, 2300 RA Leiden, The Netherlands}
\author{Dimitrios Krommydas}
\affiliation{Instituut-Lorentz, Universiteit Leiden, P.O. Box 9506, 2300 RA Leiden, The Netherlands}
\author{Luca Giomi}
\email{giomi@lorentz.leidenuniv.nl}
\affiliation{Instituut-Lorentz, Universiteit Leiden, P.O. Box 9506, 2300 RA Leiden, The Netherlands}

\date{\today}

\begin{abstract}	
Quasi-long ranged order is the hallmark of two-dimensional liquid crystals. At equilibrium, this property implies that the correlation function of the local orientational order parameter decays with distance as a power law: i.e. $\sim|\bm{r}|^{-\eta_{p}}$, with $\eta_{p}$ a temperature-dependent exponent. While in general non-universal, $\eta_{p}=1/4$ universally at the Berezinskii-Kosterlitz-Thouless  transition, where orientational order is lost because of the unbinding of disclinations. Motivated by recent experimental findings of liquid crystal order in confluent cell monolayers, here we demonstrate that, in {\em active} liquid crystals, the notion of quasi-long ranged order fundamentally differs from its equilibrium counterpart and is ultimately dictated by the interplay between translational and orientational dynamics. As a consequence, the exponent $\eta_{p}$ is allowed to vary in the range $0<\eta_{p}\le 2$, with the upper bound corresponding to the isotropic phase. Our theoretical predictions are supported by a survey of recent experimental data, reflecting a wide variety of different realization of orientational order in two dimensions. 
\end{abstract}

\maketitle

The notion of {\em order} in condensed matter differs greatly in two and three dimensions~\cite{Chaikin1995}. Two-dimensional crystals, for instance, are characterized by a power-law decaying positional correlation function -- i.e. $\langle \rho(\bm{r})\rho(\bm{0})\rangle \sim |\bm{r}|^{-\eta_{T}}$, with $\rho$ the density of the atoms and $\eta_{T}$ a positive non-universal temperature-dependent exponent -- in stark contrast with three-dimensional crystals, where the same function converges to a finite limit at long distances~\cite{Halperin1978,Nelson1979}. In two-dimensional liquid crystals, an analogous example of power-law decaying order concerns the complex function $\psi_{p}=e^{ip\vartheta}$, with $p$ an integer reflecting the symmetry of the ordered phase ($p=1$ polar order, $p=2$ nematic order, etc.) and $\vartheta$ the local orientation of the mesogens~\cite{Degennes1993}. As in the case of translational order $\langle \psi_{p}^{*}(\bm{r})\psi_{p}(\bm{0})\rangle \sim |\bm{r}|^{-\eta_{p}}$, with $\eta_{p}$ another non-universal exponent~\cite{Kosterlitz1972,Kosterlitz1973}. This behavior is also reflected by the orientational order parameter $\Psi_{p}=\langle \psi_{p} \rangle_{\ell}$, with $\langle\cdots\rangle_{\ell}$ an ensemble average over a domain of size $\ell$ (see, e.g., Refs.~\cite{Giomi2022a,Giomi2022b,Armengol2023}), whose magnitude scales like $|\Psi_{p}|\sim\ell^{-\eta_{p}/2}$ and vanishes at the macroscopic scale. The latter motivates the terminology quasi long-ranged order (QLRO), used to distinguish two-dimensional crystals and liquid crystals from their three-dimensional counterparts, where translational and orientational order are long-ranged and the order parameter is independent of the scale $\ell$ at which is probed (provided this is much larger than any microscopic length scale) and finite in the infinite system size limit. 

At thermal equilibrium, the exponents $\eta_{T}$ and $\eta_{p}$ increase linearly with temperature and attain their maximum at the solid-liquid and anisotropic-isotropic phase transitions respectively, where QLRO becomes unstable to the unbinding of topological defects~\cite{Kosterlitz1972,Kosterlitz1973,Halperin1978,Nelson1979}. For orientational order, in particular
\begin{equation}\label{eq:eq1}
\eta_{p}=\frac{p^{2}k_{B}T}{2\pi K}\;,
\end{equation}
where $K$ is the orientational stiffness of the associated $p-$atic phase~\cite{Giomi2022a,Giomi2022b}. At the Berezinskii-Kosterlitz-Thouless transition $k_{B}T_{\rm c}=\pi K/(2p^{2})$ and pairs of $\pm 1/p$ topological defects are entropically favoured to unbind. Thus 
\begin{equation}\label{eq:eq2}
\eta_{p} \le 1/4\;,
\end{equation}
in the ordered phase, whereas the equal sign holds exclusively at criticality~\cite{Kosterlitz1972,Kosterlitz1973}.

While these concepts are firmly rooted in the statistical physics of {\em passive} two-dimensional matter at equilibrium, recent observations of the phase behavior of two-dimensional {\em active} liquid crystals~\cite{Marchetti2013,Ramaswamy2017,Doostmohammadi2018,You2019,carenza_review,Shankar2022} -- i.e. liquid crystals whose building blocks can autonomously move and perform mechanical work -- have indicated that Eq.~\eqref{eq:eq2} could be systematically violated in this context, while simultaneously creating the demand for a generalization of the notion of QLRO beyond the classic assumptions of equilibrium phase transitions. Specifically, recent experiments on confluent layers of Madin-Darby canine kidney (MDCK) epithelial cells have revealed the existence of both nematic (i.e. $p=2$) and hexatic (i.e. $p=6$) order, with the former being dominant at large and the latter at short length scales~\cite{Armengol2023,Eckert2023,Armengol2022,Krommydas2023}. For both classes of $p-$atic order, the measured value of $\eta_{p}$ is significantly larger than the classically predicted upper bound of $1/4$, with $\eta_2=1.02 \pm 0.04$ and $\eta_6=1.50 \pm 0.06$. These findings have raised questions about the nature of orientational order in active liquid crystals and whether the notion of QLRO is still the most suited to describe it.  

In this article, we address this problem using a combination of numerical simulations, analytical work and data from the existing experimental literature. We demonstrate that, once the constraint of equilibrium is lifted, Eq.~\eqref{eq:eq2} no longer holds and $\eta_{p}$ is allowed to vary in the range $0 < \eta_{p} \le 2$, with the upper bound corresponding to the isotropic phase. For the most established case of active nematics, using an effective model of active orientational dynamics, we show that these two-dimensional active liquid crystals approach the isotropic state from a continuous spectrum of progressively disordered configurations. Remarkably, the exponent $\eta_{2}$ resulting from this mechanism is, in first approximation, independent of the magnitude of the active stress fuelling the flow. These predictions are then confirmed by a numerical integration of the hydrodynamic equations of active nematics in the regime of ``active turbulence''~\cite{Giomi2015, Doostmohammadi2017, Alert2020, Carenza2020_1, Carenza2020_2, Martinez2021, Alert2022}. Finally, we present a survey of experimental realizations of two-dimensional active liquid crystals featuring {\em in vitro} mixtures of microtubules and kinesin~\cite{Martinez2021,Pearce2021,Serra2023}, actomyosin fluids~\cite{Kumar2018}, suspensions of planktonic bacteria~\cite{Wensink2012,Li2019,Copenhagen2021} and eukaryotic cell cultures~\cite{Blanch2013,Armengol2023}. For all these systems, we find that $\eta_{p}$ is larger than $1/4$, therefore confirming that two-dimensional active liquid crystals comprise in fact a distinct class of partially ordered systems, intermediate between quasi-long-ranged and short-ranged order matter. 

Let us consider an incompressible active nematic liquid crystal, whose configuration is described by the director field $\bm{n}=\cos\theta\,\bm{e}_{x}+\sin\theta\,\bm{e}_{y}$ and the divergenceless velocity field $\bm{v}$ (i.e. $\nabla\cdot\bm{v}=0$), dynamics is governed by the following hydrodynamic equations:
\begin{subequations}\label{eq:eq3}
\begin{gather}
\frac{D\theta}{Dt} = D_{\rm r}\nabla^{2}\theta+\frac{\omega}{2}-\lambda(u_{xx}\sin 2\theta-u_{xy}\cos 2\theta)\;,\\
\rho\,\frac{D\bm{v}}{Dt} = \nabla\cdot\left(\Sp+\Sa\right)\;,
\end{gather}
\end{subequations}
with $D/Dt=\partial_{t}+\bm{v}\cdot\nabla$ the material derivative. In Eq.~(\ref{eq:eq3}a), $D_{\rm r}=K/\gamma$, with $K$ the orientational stiffness of the nematic phase and $\gamma$ the rotational viscosity, is a rotational diffusion coefficient, $\omega=\partial_{x}v_{y}-\partial_{y}v_{x}$ the vorticity of the flow,  $u_{ij}=(\partial_{i}v_{j}+\partial_{j}v_{i})/2$ the strain-rate tensor and $\lambda$ the flow alignment parameter. In Eq.~(\ref{eq:eq3}b), on the other hand, $\rho$ is the density of the fluid and the tensors $\Sp$ and $\Sa$ embody the passive and active stresses sourcing the flow respectively. The former, is given by $\sigma_{ij}^{({\rm p})}=-P\delta_{ij}+2\eta u_{ij}+(K/2)\epsilon_{ij}\nabla^{2}\theta-K\partial_{i}\theta\partial_{j}\theta$, with $P$ the pressure, $\eta$ the shear viscosity and $\epsilon_{ij}$ the two-dimensional antisymmetric tensor. The active stress, by contrast, is given by $\sigma^{({\rm a})}_{ij}=\alpha(n_{i}n_{j}-\delta_{ij}/2)$, with $\alpha$ the magnitude of the contractile (for $\alpha>0$) or extensile (for $\alpha<0$) stresses exerted by active nematogens~\citep{Hatwalne2004}. 

Now, the active flow causes a distortion of the nematic director, which, in turn, is counterbalanced by the entropic elasticity of the nematic phase, which acts towards restoring a uniform orientation throughout the system. This gives rise to coherent structures, such as bands and vortices, whose typical size, $\ell_{\rm a} = \sqrt{K/|\alpha|}$, and turnover time, $\tau_{\rm a}=\eta/\alpha$, reflect the interplay between the active and passive torques at play~\cite{Giomi2015,Hoffmann2022}. When $\ell_{\rm a}$ is much smaller than the system size, in particular, the resulting hydrodynamic state, often referred to as {\em active turbulence}, is characterized by the dynamical equilibrium between the unbinding and annihilation of topological defects, so that their mean number is conserved and proportional to $\ell_{\rm a}^{-2} \sim \alpha$. Such a regime, is by far the most commonly observed in {\em all} experimental realizations of active nematics. Furthermore, the presence of topological defects per se suggests that Eq.~\eqref{eq:eq2}, expressing the stability of the ordered phase with respect to defect unbinding, is in fact {\em violated}. An alternative upper bound for the exponent $\eta_{p}$ can instead be readily found by noticing that, by virtue of the central limit theorem, averaging an arbitrary number $N$ of uniformly distributed orientations yields $|\Psi_{p}| \sim 1/\sqrt{N}$. As $N \sim \ell^{d}$ in a $d-$dimensional space, one has that $|\Psi_{p}| \sim \ell^{-1}$ for $d=2$, hence $\eta_{p} = 2$ in the maximally disordered configuration.

\begin{figure}[t!]
\includegraphics[width=1.0\columnwidth]{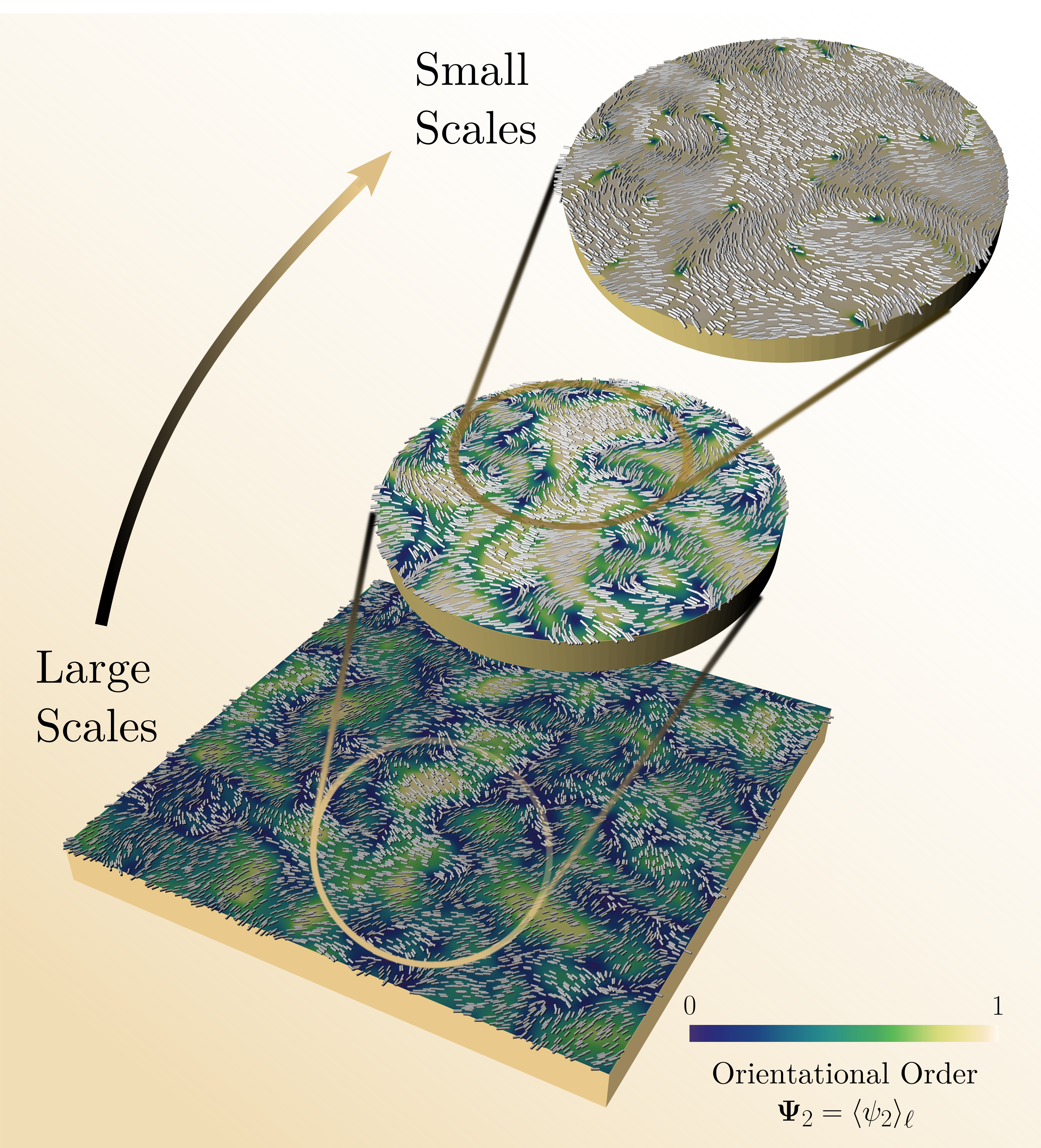}
\caption{\textbf{Coarse-graining in turbulent active nematics.} Graphical representation of the coarse-graining procedure in a turbulent active nematics. When resolved at the smallest scale magnitude of orientational order is large everywhere except in proximity of topological defects. At larger coarse-graining scales the field appears smoother, and its magnitude decreases.}
\label{Fig:1}
\end{figure}

To make progress, in the following we present a simple fluctuating hydrodynamics calculation built upon the mean-field theory of active turbulence introduced in Ref.~\cite{Giomi2015}. To this end, we set $\lambda=0$ without loss of generality (see e.g. \cite{Giomi2015,Alert2020}), and coarse-grain Eq.~(\ref{eq:eq3}a) over the length scale $\ell_{\rm a}$ and time scale $\tau_{\rm a}$, so that $\overline{\theta}$ is the average orientation of the nematic director at the typical length and time scale of the vortices. Thus, averaging both sides of Eq.~(\ref{eq:eq3}a) gives
\begin{equation}\label{eq:eq4}
\partial_{t} \overline{\theta} = (D_{\rm r} + D_{\rm a}) \nabla^{2} \overline{\theta} + \frac{\overline{\omega}}{2}\;,
\end{equation}
where, borrowing a standard approximation from eddy diffusion (see e.g. Ref.~\cite{Frisch1995}), we have set $\overline{\bm{v}\cdot\nabla\theta} \approx -D_{\rm a}\nabla^{2}\overline{\theta}$. The quantity $D_{\rm a}$ is a rotational analog of eddy diffusivity and expresses the additional contribution to orientational diffusion resulting from the advection by the active turbulent flow. From dimensional analysis $D_{\rm a} \sim \ell_{\rm a}^{2}/\tau_{\rm a}=K/\eta$ and is thus {\em independent} of the active stress $\alpha$. Interestingly, the resulting effective rotational diffusion coefficient has the same structure of that found in the Stokes regime of a passive $p-$atic liquid crystal, where spatial variations of the local orientation are assumed small~\cite{Giomi2022b}. The same coarse-graining procedure could now be applied to Eq.~(\ref{eq:eq3}b), in order to compute the coarse-grained vorticity $\overline{\omega}$. Taking advantage of the loss of coherence of the vortices at distances much larger than $\ell_{\rm a}$, and times much longer than $\tau_{\rm a}$, 
the coarse-grained vorticity $\overline{\omega}$ can be factually treated as an {\em independent} random field, whose correlation function in Fourier space and real time is given by
\begin{equation}\label{eq:eq5}
\langle \overline{\omega}(\bm{q},t)\overline{\omega}(\bm{q}',t') \rangle = (2\pi)^{2} \langle|\overline{\omega}(\bm{q})|^{2}\rangle\delta(\bm{q}+\bm{q}')\delta(t-t')\;,	
\end{equation}
with $\langle |\overline{\omega}(\bm{q})|^{2} \rangle$ the static spectral density of the vorticity field. A mean-field approximation of this function was computed in Ref.~\cite{Giomi2015} and is given by
\begin{equation}
\langle|\overline{\omega}({\bm{q}})|^{2}\rangle = 2\varsigma D_{\rm a}  e^{-\frac{\kappa^{2}}{2}}\left[I_{0}\left(\frac{\kappa^{2}}{2}\right)-I_{1}\left(\frac{\kappa^{2}}{2}\right)\right]\;,
\end{equation}
where $\kappa=\ell_{\rm a}|\bm{q}|$, $\varsigma$ is a numerical pre-factor reflecting the small-scale structure of the vortices and $I_{n}$, with $n=0,\,1$, modified Bessel functions of the first kind. At length scales much larger than the average vortex size, $\kappa \ll 1$ and $\langle|\overline{\omega}({\bm{q}})|^{2}\rangle \to 2\varsigma D_{\rm a}$. The latter, together with Eqs.~\eqref{eq:eq4} and \eqref{eq:eq5}, implies that the coarse-grained orientation $\overline{\theta}$ evolves in time according to a Langevin equation formally identical to that governing orientational equilibrium fluctuations at equilibrium, with $\varsigma D_{\rm a} \leftrightarrow k_{B}T/\gamma$. Using standard manipulations (see e.g. Ref.~\cite{Giomi2022b}) one can then integrate Eq.~\eqref{eq:eq4} and compute the orientational correlation function $\langle \psi_{2}^{*}(\bm{r})\psi_{2}(\bm{0})\rangle = \lim_{t\to\infty}\langle e^{2i[\overline{\theta}(\bm{0},t)-\overline{\theta}(\bm{r},t)]}\rangle$, from which
\begin{equation}\label{eq:eq7}
\eta_{2} 
= \frac{\varsigma}{2\pi} \left(1+\frac{D_{\rm r}}{D_{\rm a}}\right)^{-1}\;.	
\end{equation}
Once again we stress that, remarkably, the exponent $\eta_{2}$ is independent of the active stress $\alpha$ up to logarithmic corrections associated with the cut-off of the effective field theory underlying Eqs.~\eqref{eq:eq4} and \eqref{eq:eq5} at the length scale $\ell_{\rm a}$. Even more importantly, the exponent $\eta_{2}$ is not subject to upper bounds other than $\eta_{2}=2$, thereby providing a unique signature of QLRO in active liquid crystals.

\begin{figure}[t!]
\includegraphics[width=1\columnwidth ]{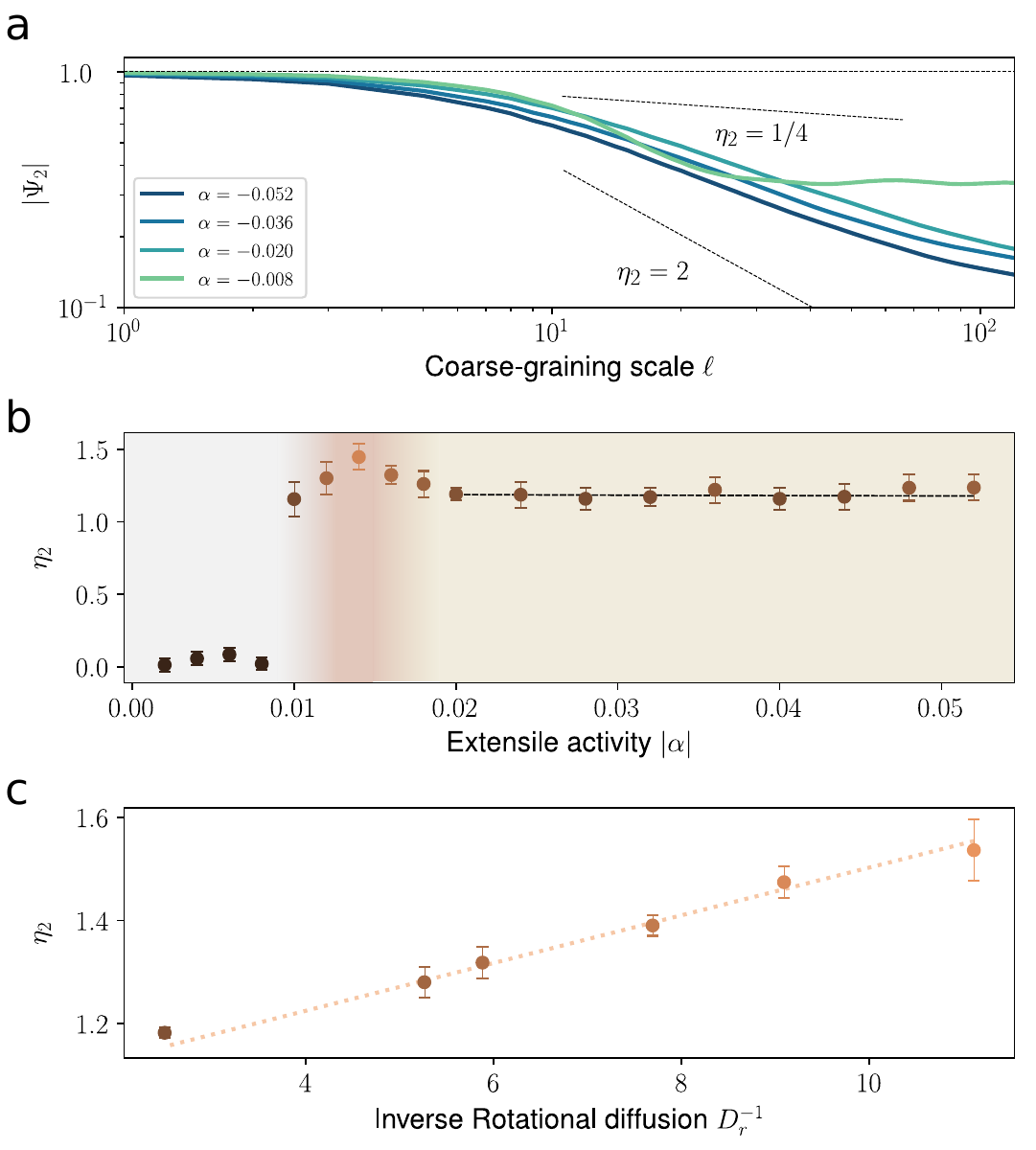}
\caption{\textbf{Exponent $\eta_{2}$ in turbulent active nematics} (a) Scale-dependent nematic order parameter $|\Psi_{2}|\sim\ell^{-\eta_{2}/2}$ in the active turbulence regime for various magnitudes of the active stress $\alpha$. 
At low activity $\alpha=0.008$ (light green curve) the system is in its quiescent state and the dynamics is dominated by equilibrium effects, leading $|\Psi_{2}|$ to plateau at larger scales.
For all other $\alpha$ values beyond the active turbulence transition, the exponent $\eta_{2}$ is larger than its equilibrium upper bound. 
(b) Exponent $\eta_2$ versus extensile activity $|\alpha|$. (c) Exponent $\eta_2$ versus the inverse of the inverse rotational diffusion coefficient $D_{\rm r}^{-1}$. All data have been generated from a Lattice-Boltzmann integration of Eqs.~\eqref{eq:eq3}; see Ref.~\cite{SI} for details and an explanation of the Lattice-Boltzmann units used in the plots.}
\label{Fig:2}
\end{figure}

\begin{figure*}[t!]
\includegraphics[width=0.8 \textwidth ]{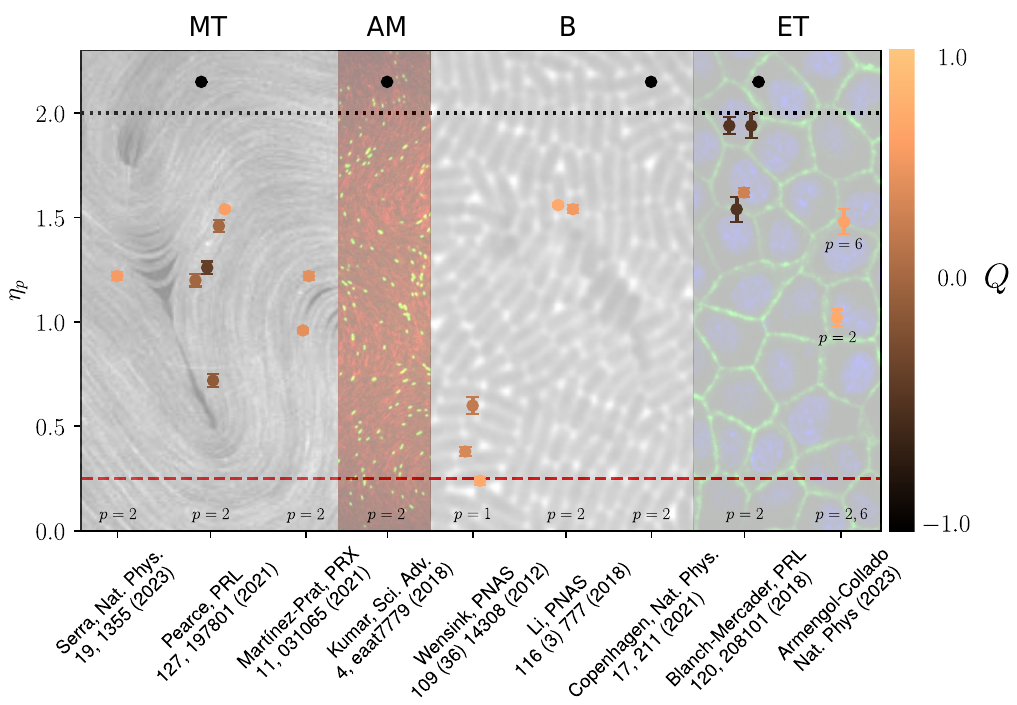}
\caption{\textbf{Exponent $\eta_{p}$ in different experimental realizations of active systems.} The data are grouped in four categories based on different realizations of active liquid crystals. From left to right: microtubules suspensions (MT), actomyosin suspensions (AM), bacterial cultures (B), and epithelial tissues (ET). The $p-$atic degree of liquid crystal order has been reported for each class of data. Notice that in Ref.~\cite{Armengol2023} both nematic  ($p=2$) and hexatic ($p=6$) order were reported. Both exponents $\eta_2$ and $\eta_6$ have been plotted and labeled accordingly. The red dashed horizontal line marks the classical upper bound for the exponent $\eta_p=1/4$. Data points are color-coded based on the {\em quality factor} $Q$, which takes the value $Q=1$ ($Q=-1$) for an exact power (exponential) law. For $Q \approx 0$, it is not possible to discriminate between a power-law and an exponential behavior (see Ref.~\cite{SI} for details). Data whose quality factor lies in the range $-1<Q<-0.5$ are reported as an overflow, above the black dotted line, and the corresponding exponent $\eta_{p}$ is considered unphysical.}
\label{Fig:3}
\end{figure*}

To assess the significance of these simple analytical predictions, we numerically integrate Eqs.~\eqref{eq:eq3}, with none of the simplifying assumptions and approximations leading to Eq.~\eqref{eq:eq7}, via a hybrid lattice Boltzmann method (see Ref.~\cite{SI} for details). As very well documented in the literature, active nematic liquid crystals under confinement exhibit a transition from a quiescent state, where the system is stationary and uniformly oriented, to a regime of active turbulence, where activity-induced whirling flows span across the plane, leading to the proliferation of defects, which nucleate and annihilate at a constant rate. In order to investigate how the proliferation of defects affects orientational order, we compute the coarse-grained order parameter $\Psi_{2}=\Psi_{2}(\ell)$ by averaging the nematic director within a disk of radius $\ell$~\footnote{The coarse-grained order field ${\Psi}_{p}$ measures the degree of alignment of the system constituents at position $\bm{r}$  by averaging the local order parameter $\psi_p$ in a disk $D_{\ell}(\bm{r})$ of radius $\ell$ centered at $\bm{r}$, that is: $ {\Psi}_{p}(\bm{r},\ell) = \int_{D_{\ell}(\bm{r})} {\rm d}\bm{r}'/(\pi\ell^{2})\,\psi_{p}(\bm{r}')$.} (see Fig.~\ref{Fig:1}). We perform this average over uncorrelated configurations sampled at different times and we repeat the procedure for increasingly large $\ell$ values. A plot of $|\Psi_{2}|$ versus $\ell$ is reported in Fig.~\ref{Fig:2}a for the specific case of extensile activity (i.e. $\alpha<0$). In the quiescent regime, corresponding to small $\alpha$ values (light green curve), $\Psi_{2}$ decays slowly with an exponent $\eta_{2}<1/4$, consistently with the equilibrium picture. This scaling behavior, however, changes dramatically once $\alpha$ is sufficiently large for active turbulence to develop (dark blue curves). In this regime the order parameter crosses over from slowly to quickly decaying, with an exponent $\eta_{2}$ considerably larger than $1/4$. As expected, the crossover occurs at $\ell\approx\ell_{\rm a}$, thus at a length scale that becomes increasingly shorter as the active stress is increased. A systematic analysis of the magnitude of $\eta_{2}$ at length scales larger than $\ell_{\rm a}$ is reported in Fig.~\ref{Fig:2}b and shows that, after the transition from quiescence to active turbulence, $\eta_{2}$ is consistently larger than one. Furthermore, well in the active turbulence regime, $\eta_{2}$ is roughly insensitive to the the active stress $\alpha$, in agreement with our analytical predictions. To test these further, we select different values of the ratio $D_{\rm r}/D_{\rm a}\sim\eta/\gamma$ by varying the orientational viscosity $\gamma$, while setting $|\alpha|$ large enough to ensure a fully developed active turbulence. Consistently with our predictions, the exponent $\eta_{2}$ depends linearly on $D_{\rm r}^{-1}$ (see Fig.~\ref{Fig:2}c).

To conclude our analysis of QLRO in active liquid crystals, in Fig.~\ref{Fig:3} we present a survey of experimental realizations with different $p-$atic symmetry with $p=1,\,2$ and $6$~\cite{Martinez2021,Pearce2021,Serra2023,Kumar2018,Wensink2012,Li2019,Copenhagen2021,Blanch2013,Armengol2023}. These are organized in four classes: i.e. microtubules and actomyosin suspensions, bacterial cultures and epithelial cell layers. The former two classes provide a quintessential example of cytoskeletal active nematics, while elongated bacteria, such as the strains of {\em B. subtilis}, {\em E. coli} and {\em M. xanthus} used in Refs.~\cite{Wensink2012,Li2019,Copenhagen2021} feature either polar (i.e. $p=1$) or nematic (i.e. $p=2$) order, depending on whether the cells express a planktonic (i.e. swimming) or sessile (i.e. dividing) phenotype. Among epithelial layers, on the other hand, it is possible to observe realizations of both hexatic (i.e. $p=6$) and nematic order, depending on the cell type and the observed range of scales. For the MDCK cells investigated in Refs.~\cite{Armengol2023} and included in this survey, both types of orientational order are present, with the former being dominant at the small and the latter at the large scale. Noticeably, among all the analyzed experimental data, and irrespective of the specific $p-$atic order, the measured exponent $\eta_{p}$ is {\em systematically larger} than $1/4$ and often attains magnitudes which are seven times larger than that of the equilibrium upper bound (i.e. $\eta_{p}=1.75$), while still remaining below the threshold of isotropy (i.e. $\eta_{p}=2$). Because of the limited range of the data, the power law scaling is often difficult to distinguish from an exponential decay. To overcome this limitation, the data in Fig.~\ref{Fig:3} have been color-coded by a {\em quality factor} $Q$, such that $Q=1$ ($Q=-1$) corresponds an exact power (exponential) law (see Ref.~\cite{SI} for details).

In summary, we have investigated the nature of orientational order in active liquid crystals, using numerical and experimental data, as well as  analytical arguments. We demonstrated that, while QLRO order is present in these systems, it decays with distance faster than at equilibrium. The latter is quantified via the exponent $\eta_{p}$, dictating the rate of power-law decay of the orientational correlation function and of the scale-dependent order parameter. In both passive and active liquid crystals, the exponent $\eta_{p}$ is non-universal, but depends upon the system temperature and material properties. Yet, while at equilibrium $\eta_{p}=1/4$ universally at the Berezinskii-Kosterlitz-Thouless transition, where orientational order is lost due to disclination unbinding, this upper bound is generally violated in active liquid crystals and $\eta_{p}$ can take arbitrary values in the range $0<\eta_{p} \le 2$, with the upper bound corresponding to the isotropic phase. Whereas the notion of QLRO may still be appropriate to classify order in active liquid crystals, our analysis shows that this class of non-equilibrium fluids in fundamentally less ordered than their passive counterpart, is manifest from the abundance of topological defects. 
Another important observation regards the range of validity of the active liquid crystal theory. Indeed, this has been developed as an extension of the de Gennes' hydrodynamic theory for liquid crystals and, therefore, inherently implies the existence of an ordered phase.
At the same time, the good agreement between theory and experiments, demonstrated in the last decade in the literature, highlights the robustness of the existing theoretical framework that exceeds expectations
and encourages even broader applications of the paradigm of active matter in the near future.

\acknowledgements

The authors would like to express their profound gratitude to Berta Martinez-Praz, Francesc Sagues-Mestre, Nitin Kumar, Margaret Gardel, Mattia Serra, Linnea Lemma, Zvonimir Dogic, Victor Yashunsky, He Li and Hepeng Zhang for sharing the experimental data that made this analysis possible. This work is supported by the ERC-CoG grant HexaTissue and by Netherlands Organization for Scientific Research (NWO/OCW). The computational work was carried out on the Dutch national e-infrastructure with the support of SURF through the Grant 2021.028 for computational time. The authors acknowledge Ludwig Hoffmann for fruitful discussions.

\bibliography{Biblio.bib}
\end{document}


\title{Quasi long-ranged order in two-dimensional active liquid crystals}

\author{Livio Nicola Carenza}
\affiliation{Instituut-Lorentz, Universiteit Leiden, P.O. Box 9506, 2300 RA Leiden, The Netherlands}

\author{Josep-Maria Armengol-Collado}
\affiliation{Instituut-Lorentz, Universiteit Leiden, P.O. Box 9506, 2300 RA Leiden, The Netherlands}

\author{Dimitrios Krommydas}
\affiliation{Instituut-Lorentz, Universiteit Leiden, P.O. Box 9506, 2300 RA Leiden, The Netherlands}

\author{Luca Giomi}
\email{giomi@lorentz.leidenuniv.nl}
\affiliation{Instituut-Lorentz, Universiteit Leiden, P.O. Box 9506, 2300 RA Leiden, The Netherlands}

\maketitle

\section{Numerical Model for Active Nematics and Computational Details}
In this Section we present the full set of hydrodynamics equations used to numerically investigate the ordering properties of a turbulent active nematics.
The nematohydrodynamics of a $2D$ active gel can be described in terms of the $Q-$tensor and the velocity field $\bm{v}$. The former is a traceless and symmetric tensor encoding the local orientational features of the active liquid crystal. In $2D$ this can be written as $\bm{Q} =  S (\bm{n}\bm{n} - \bm{I}/2)$, where $S$ is a measure of the magnitude of nematic order, $\bm{n}= \cos{\theta}\, \bm{e}_x + \sin{\theta}\, \bm{e}_y$ is the director field defining the local direction of alignment, and $\bm{I}$ the identity tensor.
The evolution of the $\bm{Q}$-tensor is ruled by the Beris-Edwards equation:
\begin{equation}
(\partial_t + \bm{v}\cdot \nabla) \mathbf{Q} - \mathbf{S}(\nabla \bm{v},\mathbf{Q}) = \gamma^{-1} \mathbf{H} \, .  
\label{eqn:S1}
\end{equation}
Here, the strain-rotational derivative
\begin{equation}
\mathbf{S}(\nabla \bm{v},\mathbf{Q}) = (\lambda \mathbf{D} + \mathbf{\Omega})(\mathbf{Q}+\mathbf{I}/2) + (\mathbf{Q}+\mathbf{I}/2)(\lambda \mathbf{D}  - \mathbf{\Omega}) - 2 \lambda (\mathbf{Q}+\mathbf{I}/2) \rm{Tr} (\mathbf{Q} \nabla \mathbf{v}) \; ,
\label{eqn:S2}
\end{equation}
couples the liquid crystal dynamics to that of the underlying fluid, where $\bm{D}= (\nabla \bm{v} + \nabla \bm{v}^T)/2$ is the strain rate tensor, $\bm{\Omega} = (\nabla \bm{v} - \nabla \bm{v}^T)/2$ is the vorticity tensor, and $\lambda$ is the flow-alignment parameter.
The term on the right-hand side drives the relaxation of the liquid crystal towards its ground state. In particular, $\gamma$ is the rotational viscosity and $ \bm{H} = - \frac{\delta \mathcal{F}}{\delta \bm{Q}} + \frac{1}{2} {\rm Tr}\left(\frac{\delta \mathcal{F}}{\delta \bm{Q}} \right) \bm{I}$ is the molecular field derived from the Landau-De Gennes free energy 
\begin{equation}
\mathcal{F} = \int d \bm{r} \left[ A_0 \left( \frac{1}{2} \left(1 - \frac{\chi}{3} \right)\mathbf{Q}^2 -  \frac{\chi}{3} \mathbf{Q}^3 +  \frac{\chi}{4} \mathbf{Q}^4 \right) + \frac{K}{2} (\nabla  \mathbf{Q})^2  \right]
\label{eqn:S3}
\end{equation}
where $A_0$ and $K$ are respectively the bulk and elastic constant and $\chi$ is a temperature-like parameter controlling the isotropic-nematic transition, occurring at  $\chi \geq 2.7$.
The evolution of the flow field $\bm{v}$ is governed by the incompressible Navier-Stokes equation:
\begin{equation}
\rho (\partial_t + \bm{v} \cdot \nabla ) \bm{v} = -\nabla p + \nabla \cdot \left[ \bm{\sigma}^{pass} + \bm{\sigma}^{act} \right], 
\label{eqn:S4}
\end{equation}
where $\rho$ is the constant fluid density, $p$ is the ideal fluid pressure enforcing fluid incompressibility ($\nabla \cdot \bm{v} = 0$) and the stress tensor has been divided in a \emph{passive} ($\sigma^{\rm pass}$) and an \emph{active} ($\sigma^{\rm act}$) part.
The former accounts for dissipative and reactive effects and can be expressed as the sum $\sigma^{\rm pass} = \sigma^{\rm visc} + \sigma^{\rm el}$  of a viscous contribution $\sigma^{\rm visc}= 2 \eta \mathbf{D}$, with $\eta$ the fluid viscosity, and an elastic contribution
\begin{equation}
\bm{\sigma}^{\rm el} = -\lambda \bm{H} \left( \bm{Q} + \dfrac{\bm{I}}{3} \right) - \lambda  \left(\bm{Q} + \dfrac{\bm{I}}{3} \right) \bm{H} + 2\lambda {\rm Tr}(\bm{Q} \bm{H}) \left(\bm{Q} - \dfrac{\bm{I}}{3} \right)  + \bm{Q} \bm{H}  - \bm{H} \bm{Q} . 
\label{eqn:S5}
\end{equation}
The active stress is given by $\sigma^{\rm act}= \alpha \mathbf{Q}$. The constant $\alpha$, the so-called activity, is a phenomenological term which tunes the intensity of the active doping and describes extensile particles, if $\alpha<0$, or contractile ones otherwise.

Eqs.~\eqref{eqn:S1}) and~\eqref{eqn:S4} were integrated in a box of size $L=512$ with periodic boundary conditions through a hybrid lattice Boltzmann method, where the hydrodynamics is solved with a \emph{predictor-corrector} LB algorithm, while the dynamics of the $Q-$tensor is integrated with a finite-difference approach implementing a first-order upwind scheme and fourth-order accurate stencil for the computation of spacial derivatives. 

The numerical code has been parallelized by means of Message Passage Interface (MPI) by dividing the computational domain in slices and
by implementing the ghost-cell method to compute derivatives on the boundary of the computational subdomains.
Runs have been performed using $128$  CPUs for at least $10^7$ lattice Boltzmann iterations.

The model parameters in lattice units used for simulations are $A_0=0.04$, $K=0.08$, $\eta = 1.66$. We varied the activity $\alpha$ in the range $0.0, -0.065$ and the rotational viscosity $\gamma$ in the range $0.04, 0.4$.


\section{Order decay in experimental active fluids}
\begin{figure}[p]
\includegraphics[width=1.0 \textwidth ]{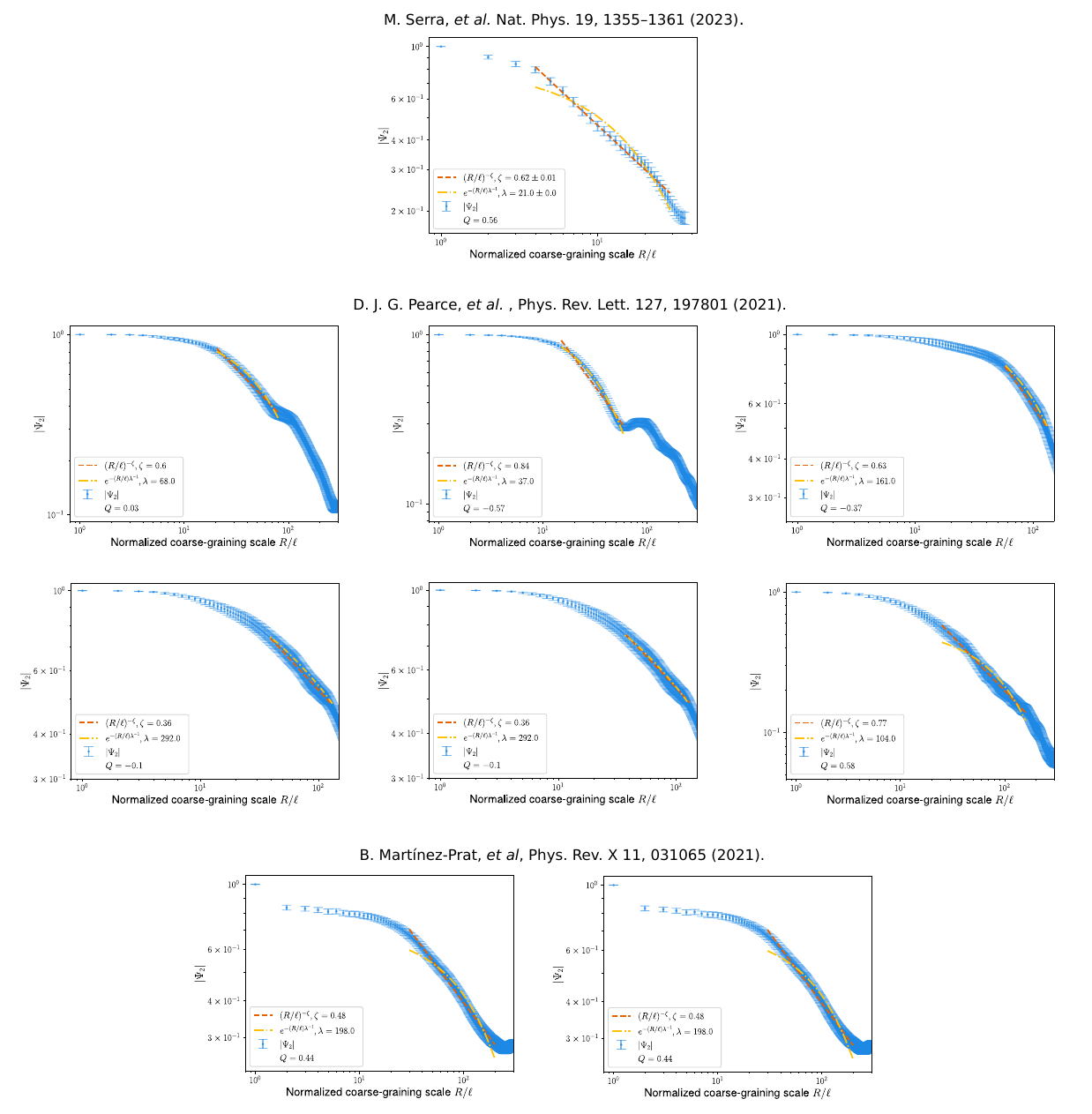}
\caption{\textbf{Orientational order in active microtubule suspensions.}}
\label{FigS1}
\end{figure}

\begin{figure}[p]
\includegraphics[width=0.38\textwidth ]{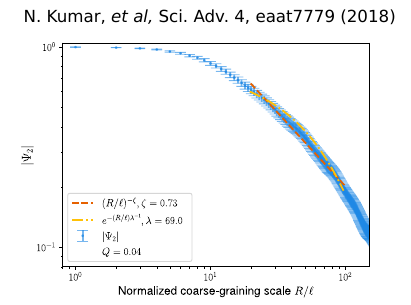}
\caption{\textbf{Orientational order in actomyosin suspensions.}}
\label{FigS2}
\end{figure}

\begin{figure}[p]
\includegraphics[width=1.0\textwidth ]{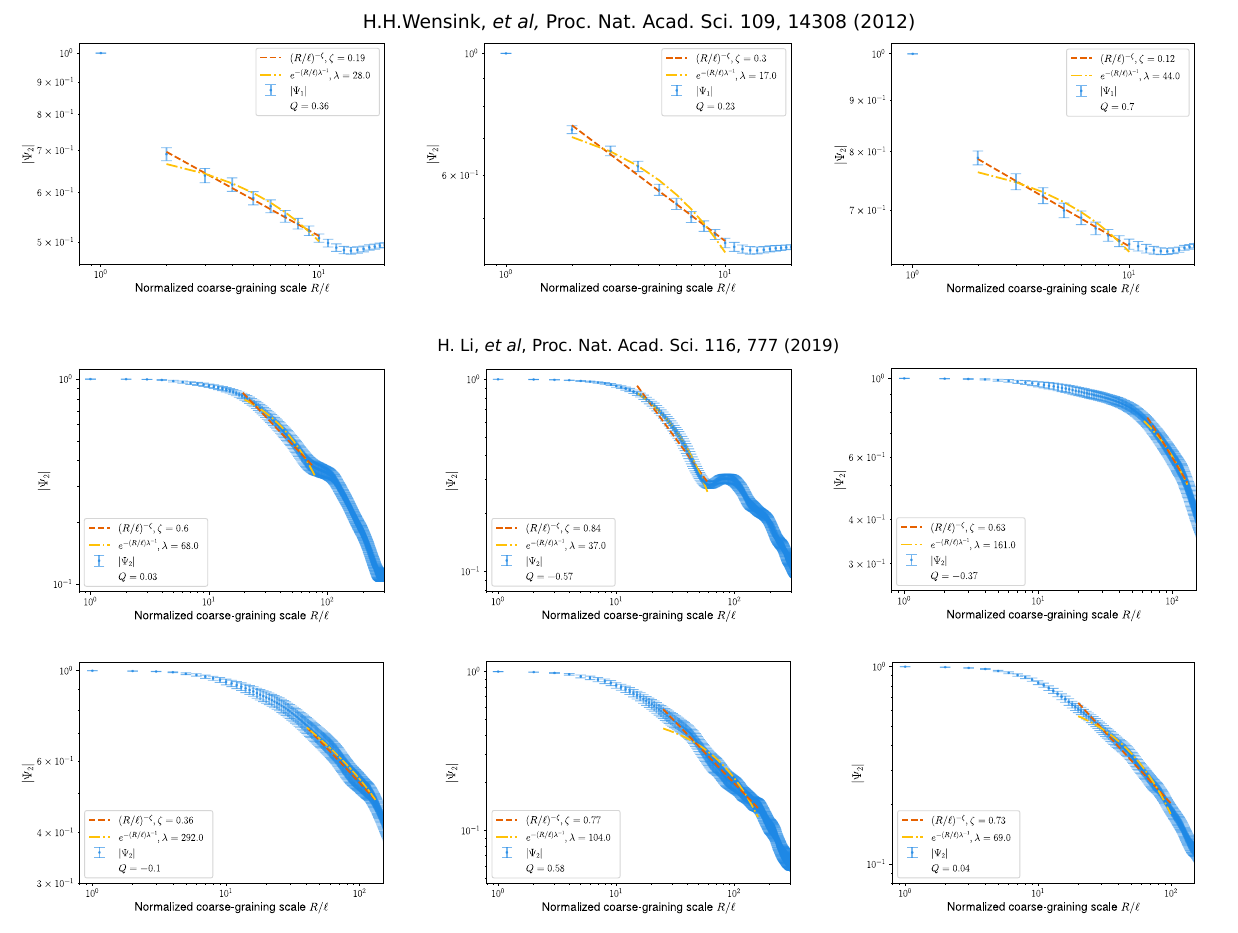}
\caption{\textbf{Orientational order in bacterial fluids.}}
\label{FigS3}
\end{figure}

\begin{figure}[p]
\includegraphics[width=1.0\textwidth ]{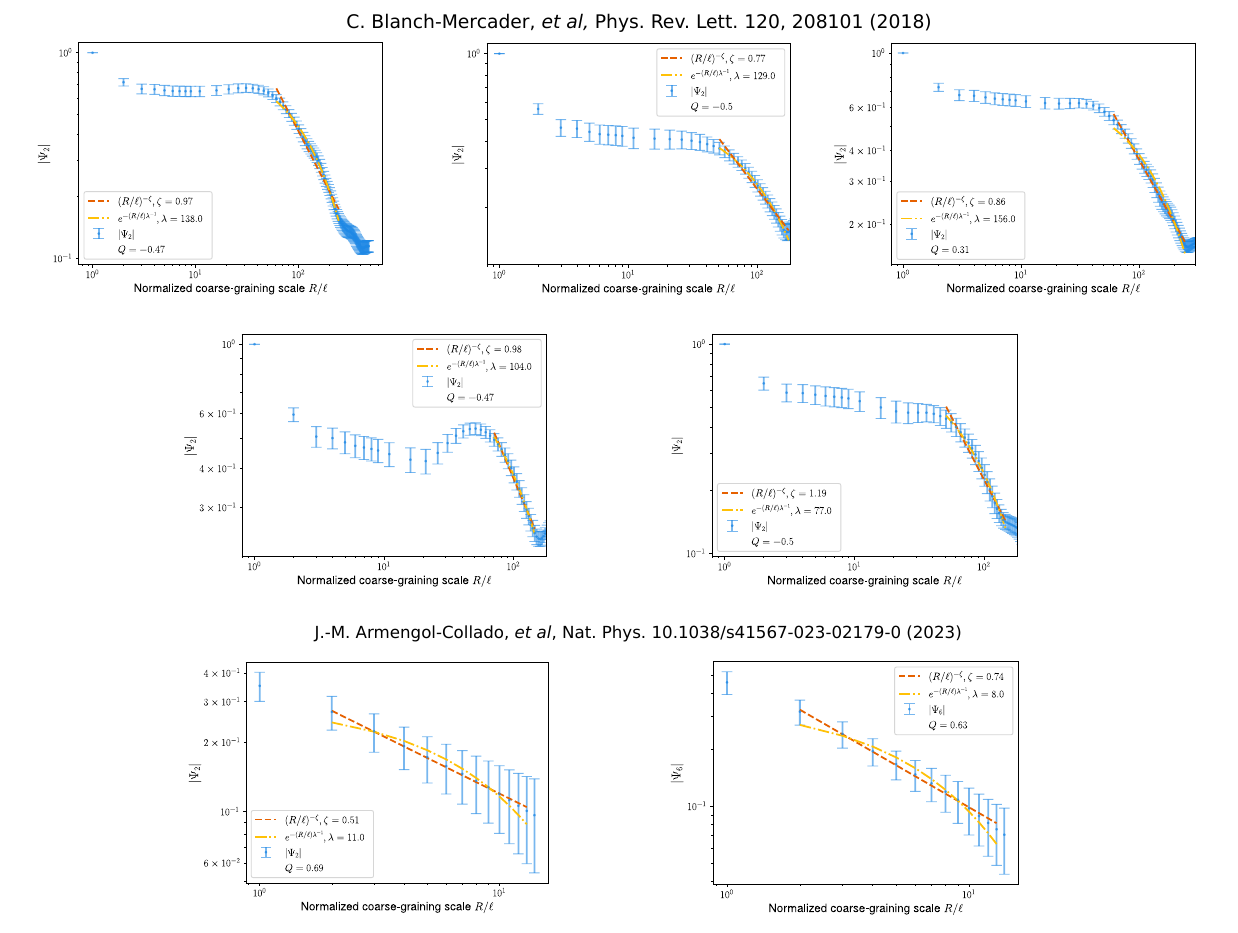}
\caption{\textbf{Orientational order in epithelial tissues.}}
\label{FigS4}
\end{figure}

Supplementary Figures~\ref{FigS1}-\ref{FigS4} show a gallery of the decay of the magnitude of the order parameter $|\Psi_p|$ at increasing coarse-graining scale in a variety of different active fluids featuring orientational order. In particular, Fig.~\ref{FigS1} shows data from references [23,25,26] for the case of active nematic microtubule suspensions; Fig.~\ref{FigS2} from reference [27] for an actomyosin fluid; Fig.~\ref{FigS3} shows data from bacterial systems both for swimming bacteria (planktonic) featuring polar symmetry ($p=1$), as in the case of Ref.~[28] and sessile featuring nematic symmetry   ($p=2$) in the case of Ref.~[29]. The orientational correlation function pertaining to the reference cited as [30] can be found in Extended Data Figure 2 of the same paper. The authors report exponential decay of the orientational correlations. Fig.~\ref{FigS4} refers to epithelial tissues. In this case, Ref.~[31] only consider the case of nematic order, while in Ref.~[9] both nematic and hexatic ($p=6$) symmetries are considered.

For each of the aforementioned cases, we started by coarse-graing the local orientational order parameter $\psi_p$ over a scale $\ell$. The details of the coarse-graining procedure are described in the Methods section of Ref.~[9]. For each coarse-graining radius, we perform an ensamble average over uncorrelated configurations to estimate the magnitude of the orientational order at different lengthscales. The results of this analysis are shown in Supplementary Figures~\ref{FigS1}-\ref{FigS4}.
We fit the data with both a power-law function $\mathcal{P}(R) = A (R/\ell)^{-\zeta}$ and a decreasing exponential $\mathcal{E}(R) = A\exp(R \lambda^{-1}/\ell)$, with $\ell$ a rescaling constant. To quantify the quality of the fit we compute the $L_2-$distance between the data points and the fit,
\begin{align*}
d_{\rm pl} = \sum_{\ell} [|\langle\Psi_p(\ell) \rangle | - \mathcal{P}(\ell)]^2 \;, \\
d_{\rm exp} = \sum_{\ell} [|\langle\Psi_p(\ell) \rangle | - \mathcal{E}(\ell)]^2 \;.
\end{align*}
and we define the quality factor 
$$
Q= \dfrac{d^{-1}_{\rm pl}-d^{-1}_{\rm exp}}{d^{-1}_{\rm pl}+d^{-1}_{\rm exp}}.
$$
The quality factor $Q \rightarrow 1$ for data distributions which decay with power law behavior; $Q \rightarrow -1$ for data distributions which decay exponentially. Finally, the case $Q \approx 0$ signals that, in the range of the fit, it is not possible to discriminate between a power-law and an exponential behavior. Fig.~3 of the manuscript shows the decay exponent $\eta_p$ estimated as $\zeta/2$ for each of the analyzed experiments. Moreover, each point is colored with the corresponding value of the quality factor $Q$. We consider datasets where $Q<-0.5$ as decaying exponentially. In this case, we regard the estimation of a decaying exponent $\eta_p$ as non physical. Therefore, we report the point as an overflow with no associated decay exponent.

